\documentstyle[12pt,epsfig,array]{article}
\def\np#1#2#3{Nucl.\ Phys.\ B#1 (19#3) #2}
\def\pl#1#2#3{Phys.\ Lett.\ #1B (19#3) #2}
\def\pr#1#2#3{Phys.\ Rev.\ D #1 (19#3) #2}
\def\prb#1#2#3{Phys.\ Rev.\ B #1 (19#3) #2}
\def\prep#1#2#3{Phys.\ Rep.\ #1 (19#3) #2}

\def\zp#1#2#3{Z.\ Phys.\ C #1 (19#3) #2}

\def\slash#1{\ooalign{$\hfil/\hfil$\crcr$#1$}}
\def\bom#1{\mbox{\boldmath$#1$}}

\def\LdL{ \Lambda\partial_\Lambda }

\def\Tr{ \mbox{Tr} }
\def\ie{ {\it i.e. }}
\newcommand{\ep}{\varepsilon}

\newcommand{\prendi}[2]
{
\left. #1 \right|_{#2}
}
\newcommand{\bracket}[2]
{
\left[ #1 \right]_{#2}
}
\newcommand{\PropI}
{
\Delta^{-1}_{\Lambda\Lambda_0}
}
\newcommand{\Prop}
{
\Delta_{\Lambda\Lambda_0}
}
\newcommand{\dd}[2]
{
\frac{\delta{#1}}{\delta{#2}}
}

\newcommand{\dede}[2]
{
{\partial{#1}\over\partial{#2}}
}

\newcommand{\F}
  {{\cal F}}

\renewcommand{\L}
  {{\cal L}}

\renewcommand{\O}
  {{\cal O}}

\renewcommand{\S}
 {\Pi}

\newcommand{\rif}[1]
  {(\ref{#1})}

\newcommand{\formula}[2]
  { \begin{equation} \label{#1} #2 \end{equation} }
\newcommand{\formulona}[2]
{
  \begin{equation} \label{#1}
    \begin{array}{lll} #2
   \end{array}
  \end{equation}
}

\newcommand{\formulaX}[1]
  { \begin{equation} #1 \end{equation} }
\newcommand{\formulonaX}[1]
{
  \begin{equation}
    \begin{array}{lll} #1 \end{array}  
  \end{equation}
}

\begin{document}
\begin{titlepage}
\renewcommand{\thefootnote}{\fnsymbol{footnote}}
\begin{flushright}
     UPRF 97-05\\
     March 1997 \\
\end{flushright}
\par \vskip 10mm
\begin{center}
{\Large \bf
Wilson renormalization group and improved perturbation theory~\footnote{
Research supported in part by MURST, Italy}}
\end{center}
\par \vskip 2mm
\begin{center}
M.\ Bonini
and  M.\ Simionato\\
\vskip 5 mm
{\it Dipartimento di Fisica, Universit\`a di Parma \\
and INFN, Gruppo Collegato di Parma, Italy}
\end{center}
\par \vskip 2mm
\begin{center} {\large \bf Abstract} \end{center}
{\small \begin{quote}
We discuss a resummed perturbation theory based on the
Wilson renormalization group.  
In this formulation the Wilsonian flowing couplings, 
which generalize the running coupling, enter directly into the loop 
expansion. 
In the case of an asymptotically free theory the flowing coupling is well
defined since the infrared Landau pole is absent. 
We show this property in the case of the $\phi^3_6$ theory.
We also extend this formulation to the QED theory and we prove that 
it is consistent with gauge invariance.\\

Pacs: 11.10.Hi, 11.15.-q, 11.15.Tk. Keywords: Wilson renormalization group, 
gauge invariance, running coupling. 
\end{quote}}
\end{titlepage}
\section{Introduction}
The formulation of quantum field theory based on the Wilson 
renormalization group \cite{Wilson}, which we will call 
$\Lambda-$RG, studies the evolution in 
the infrared  cutoff $\Lambda$ of the Wilsonian effective 
action $S(\phi;\Lambda,\Lambda_0)$, where $\Lambda_0$ is 
some ultraviolet cutoff. 
This functional is obtained by integrating out all degrees of freedom 
with momenta higher than $\Lambda$ (and lower than $\Lambda_0$) 
in the functional integral. By decreasing the scale $\Lambda$ 
and requiring that physical observables are independent of $\Lambda$ 
one obtains an evolution equation \cite{P}-\cite{others} 
for $S(\phi;\Lambda,\Lambda_0)$, 
which gives a non-perturbative definition of the theory.
In this framework one can give 
simple proofs valid at any order in perturbation theory of 
many fundamental properties such as renormalizability and
infrared finiteness. Moreover the quantum implementation of 
symmetries is very easily understood \cite{B},\cite{noi},\cite{ellw}.
Nevertheless, the practical relevance of this formulation 
may appear questionable since the $\Lambda-$RG method seems 
more complex than ordinary renormalization schemes (for instance 
dimensional regularization). The introduction of a sharp cutoff makes 
the calculation of general 
Feynman integrals more difficult. Moreover, in the case of a gauge theory,
the cutoffs break explicitly the gauge invariance and one must 
prove that the theory is invariant once the cutoffs are removed.
While the first difficulty is technical and, as we will show in 
this paper, can be avoided using a suitable cutoff function, the 
latter is a fundamental issue which must be fulfilled in order to have a 
consistent theory. In ref. \cite{B,noi}
it has been proved that, by appropriately fixing the boundary 
conditions of the $\Lambda-$RG equations, the effective action 
of a gauge theory satisfies the Slavnov-Taylor identities at 
the physical point ($\Lambda=0$ and $\Lambda_0\to\infty$). 
However this proof is inextricably linked to perturbation theory.
  
The essential advantage of the Wilson formulation is 
the fact that it provides a non-perturbative definition of the 
effective action at any scale $\Lambda$ given the action at some 
(ultraviolet) scale $\Lambda_0$.
Unfortunately the $\Lambda-$RG equation corresponds to an infinite 
system of coupled differential equations for the relevant couplings 
and the irrelevant vertices of the Wilsonian effective action and its 
solution needs some approximation.
In the last few years there have been several attempts of 
finding non-perturbative approximate solutions.
In general one truncates the space of interactions to 
few operators according to their dimension 
and/or  uses a derivative expansion \cite{trunc,Morris}. 
These truncations have been 
applied expecially to scalar theories and could be very accurate 
\cite{morrisdubna}. 
Similar methods have been applied to gauge theories \cite{truncgauge}.
In this case one has to face 
the problem of consistency between truncation and gauge invariance. 
In general one can truncate the space of interactions in such a way that 
some of the Slavnov-Taylor identities are satisfied but one can show that the
truncation is incompatible with the full set of Slavnov-Taylor identities.

In this paper we consider a recursive approach, 
first formulated in \cite{BMS}, which mimics perturbation theory analysis,
but corresponds to a resummation of higher order of the coupling 
constant.
Some remarks are useful in order to present our idea.
The $\Lambda-$RG equation allows an iterative solution which gives the 
perturbative expansion. This fact is easily seen if one introduces 
the cutoff effective action $\Gamma(\phi;\Lambda,\Lambda_0)$, which is related 
to the Wilsonian action by a Legendre transformation. 
The evolution equation for the vertices of this functional adds a loop, 
thus the vertices at loop $\ell$ are determined by the vertices at lower 
loops. 
Therefore, from the effective action at zero loop, \ie the classical 
action, one can determine this functional at any loop order.

The improved formulation is similar: a
finite number of $\Lambda-$dependent couplings, the flowing couplings, 
is sorted out.
These couplings correspond at $\Lambda=0$ to the physical couplings and
are computable at any $\Lambda$ solving a finite set of
differential equations. The remaining part of the cutoff effective 
action is obtained using recursive integral equations. 
In this way the renormalized coupling constant is replaced in the loop 
integrals by the flowing coupling at the scale given by the loop 
momenta (at least in the case of a sharp cutoff). 
This point of view is very close to the resummed
perturbation theory, in which higher order corrections reconstruct 
the running coupling constant $g(q^2)$, where $q$ is the loop momentum
\cite{renorm}. 
This resummation is well established in the large $N_f$ limit of QED, 
and it is applied as an ansatz (naive non-Abelianization procedure) 
to the QCD \cite{BB}. 
In the  latter case the one-loop running coupling constant 
diverges at the infrared Landau pole, and the 
integration over the low momenta becomes ambiguous \cite{renormalons}. 
Our improved formulation is systematic,\ie applies
equally well to the non-asymptotically free and asymptotically free
theories. In the latter case the infrared Landau pole is 
absent since the flowing coupling remains finite in all range of 
the momenta. In this paper we show this property in the case of 
the one-loop improved $\phi^3_6$ theory.

For a gauge theory it is crucial that the improved perturbation theory 
does not produce a breaking of (quantum) gauge invariance, \ie 
one has to show that the solution of 
the $\Lambda-$RG equation satisfies at $\Lambda=0$ the Slavnov-Taylor 
identities.
In this paper we consider the case of QED and we show explicitly
that the one-loop improved solution satisfies the  Ward identities 
up to negative powers of the ultraviolet cutoff $\Lambda_0$. 
As in the standard resummed perturbation theory, 
the presence of the Landau pole in the ultraviolet region implies that 
one cannot remove $\Lambda_0$ and therefore the Ward identities 
are recovered only for momenta much lower than $\Lambda_0$. 

The paper is organized as follows. In sections 2 and 3 we recall the details
of the $\Lambda-$RG formulation for the massless $\phi^4_4$ theory 
in the perturbative and in the improved case, respectively. 
In section 4 we analyze the massive $\phi^3_6$ theory 
as a pedagogical example of asymptotically free theory.
In section 5 we present the improved perturbation theory for 
QED and we compute the flowing couplings at one loop. 
In section 6 we prove that the one-loop improved formulation is
consistent with gauge invariance. In section 7 we compare 
our approach with the standard resummed perturbation theory and  section 8
contains some conclusions. 
The choice of the cutoff function and the conventions are described in 
two appendices.

\section{Remarks on perturbative \bom{\Lambda-}RG method}
In order to fix the notations, we review the usual (non-improved) 
$\Lambda-$RG formulation in the case of
massless $\phi^4$ theory in four dimensions \cite{others}.
The starting point is the evolution equation for the cutoff effective
action $\Gamma(\phi;\Lambda,\Lambda_0)$
\formula{solita.eq}
{\LdL(\Gamma(\phi;\Lambda,\Lambda_0)-\frac12\phi\cdot\PropI\phi)=
\hbar I(\phi;\Lambda,\Lambda_0)
}
where
$$
\phi\cdot\PropI\phi\equiv\int_q\phi(-q)\PropI(q)\phi(q),\quad
\int_q\equiv\int \frac{d^4 q}{(2\pi)^4}
$$
and
$$
I(\phi;\Lambda,\Lambda_0)=
-\frac12\int_q\LdL\PropI(q)
\Gamma_2^{-1}(q;\Lambda,\Lambda_0)\bar\Gamma_{\phi\phi}
(q,-q;\phi;\Lambda,\Lambda_0)\Gamma_2^{-1}(q;\Lambda,\Lambda_0).
$$
The cutoff propagator $\Delta_{\Lambda\Lambda_0}(q)$ is obtained
by multiplying the free propagator $\Delta(q)\equiv1/q^2$ with the cutoff
function $K_{\Lambda\Lambda_0}(q)$. This function cuts the frequencies below 
the infrared cutoff $\Lambda$ and above the ultraviolet cutoff $\Lambda_0$. 
The auxiliary functional $\bar\Gamma_{\phi\phi}$ depends non-linearly on the
cutoff effective action and it is defined in \cite{others}.
The physical effective action
$\Gamma(\phi)$ is extracted from $\Gamma(\phi;\Lambda,\Lambda_0)$
performing the limit $\Lambda\to0$ and $\Lambda_0\to\infty$. 
It is also convenient to introduce the functional
$$
\S(\phi;\Lambda,\Lambda_0)\equiv\Gamma(\phi;\Lambda,\Lambda_0)-
\frac12\phi\cdot(\PropI-\Delta^{-1})\phi
$$
which at the tree level coincides with $S_{cl}(\phi)$. At higher orders
the cutoff vertices $\S_{2n}(p_i;\Lambda,\Lambda_0)$ 
and $\Gamma_{2n}(p_i;\Lambda,\Lambda_0)$ are equal and become the 
physical vertices in the limit $\Lambda\to0$ and $\Lambda_0\to\infty$. 
The evolution equation \rif{solita.eq}
can be iteratively solved  by using the loop expansion
$\S^{[\ell]}=\S^{(0)}+\hbar\S^{(1)}+\ldots+\hbar^\ell
\S^{(\ell)}$. One obtains 
$$
\S^{(\ell)}(\phi;\Lambda,\Lambda_0)=\hbar\int_{\Lambda}^{\Lambda_0}
\frac{d\lambda}{\lambda} I^{(\ell-1)}(\phi;\lambda,\Lambda_0)
+\mbox{boundary conditions}.
$$
In order to specify the boundary conditions the
effective action $\S(\phi;\Lambda,\Lambda_0)$ is split into a relevant
part
$$
\S_{rel}(\phi;\Lambda,\Lambda_0)=\int_x\frac12Z(\Lambda)
\partial_\mu\phi\partial^\mu\phi+\frac12
\sigma_m(\Lambda)\phi^2+\frac1{4!}\sigma_g(\Lambda)\phi^4
$$
and an irrelevant part $\S_{irr}=\S-\S_{rel}$ (the same decomposition
also holds for the functional $ I(\phi;\Lambda,\Lambda_0)$). 
The relevant couplings are
given by
$$
Z(\Lambda)=\prendi{\partial_{p^2}\S_2}{p^2=\mu^2},\quad
\sigma_m(\Lambda)=\prendi{\S_2}{p^2=0},\quad
\sigma_g(\Lambda)=\prendi{\S_4}{p_i=\bar p_i}
$$
where $\bar p_i$ is the symmetric point, defined by 
$\bar p_i\cdot \bar p_j=\frac{\mu^2}3{(4\delta_{ij}-1)}$. The
boundary conditions for the relevant couplings are fixed at any order
at the physical scale $\Lambda=0$ by
$$
Z^{(\ell)}(0)=\delta_{\ell0},\quad
\sigma_m^{(\ell)}(0)=0,\quad
\sigma^{(\ell)}_g(0)=g\delta_{\ell0}.
$$
The boundary conditions for the irrelevant part are fixed at the 
ultraviolet scale $\Lambda=\Lambda_0$ and are trivial:
$$
\prendi{\S_{irr}^{(\ell)}}{\Lambda=\Lambda_0}=0.
$$
With these boundary conditions the recursive solution
of the evolution equation exists at any perturbative order in the physical
limit $\Lambda_0\to\infty$ and $\Lambda\to0$ for non-exceptional
configurations of external momenta \cite{others}. 

\section{Improved perturbation theory}
In this section we formulate more precisely our improved 
perturbation theory for the massless $\phi^4$ theory in four 
dimensions \cite{BMS}. We introduce  the rescaled vertices
$$
\hat\S_{2n}(p_i;\Lambda,\Lambda_0)=Z^{-n}(\Lambda)\S_{2n}
(p_i;\Lambda,\Lambda_0)
$$
which satisfy the following evolution equation 
\formula{resc.eq.n}
{(\LdL+n\frac{\dot Z}Z)\hat\S_{2n}(p_i;\Lambda,\Lambda_0)=
\hat I_{2n}(p_i;\Lambda,\Lambda_0),
} 
where the dot denotes the $\LdL$ derivative and the
$\hat I_{2n}(p_i;\Lambda,\Lambda_0)$ are the vertices 
of the functional
$I(\phi;\Lambda,\Lambda_0)$ after the rescaling. In particular one has
$$
\hat I_2(p;\Lambda,\Lambda_0)=-\frac12\int_q
\hat M(q;\Lambda,\Lambda_0)\hat\S_4(q,p,-p,-q;\Lambda,\Lambda_0)
$$
and
$$
\hat I_4(p_1,\ldots,p_4;\Lambda,\Lambda_0)=
-\frac12\int_q\hat M(q;\Lambda,\Lambda_0)
[\hat\S_6(q,p_1\dots p_4,-q;\Lambda,\Lambda_0)
$$
$$-\sum_{P}\hat\S_4(q,p_{i_1},p_{i_2},-Q;\Lambda,\Lambda_0)
\hat\Gamma_2^{-1}(Q;\Lambda,\Lambda_0)\hat\S_4(Q,p_{i_3},p_{i_4},-q;
\Lambda,\Lambda_0)],
$$
where $Q= q+p_{i_1}+p_{i_2}$, the sum is over six permutations and
the measure $\hat M$ is given by
\formula{misura}
{\hat M(q;\Lambda,\Lambda_0)\equiv\hat\Gamma_2^{-1}
(q;\Lambda,\Lambda_0)\LdL\PropI(q)\hat\Gamma_2^{-1}(q;\Lambda,\Lambda_0).
} 
To calculate the improved vertices we use an iterative
procedure starting from the improved tree level cutoff action
\formula{phi4.improved}
{\hat\Gamma^{(0)}=\frac12 \hat\phi\cdot \PropI \hat\phi+
\frac1{4!}\int_x\hat g(\Lambda)\hat\phi^4.
}
The improved perturbation theory consists in solving the evolution 
equations as for the usual perturbative expansion but in terms of  
$\hat g(\Lambda)$. This coupling will be obtained at the end of the 
iterative procedure by solving its evolution equation.

The iteration starts by inserting \rif{phi4.improved} in the r.h.s. of 
\rif{resc.eq.n}. In this way one obtains the one-loop 
relevant coupling $\hat \S_2^{[1]}|_{p=0}$ and the irrelevant
vertices $\hat\S_{2n,irr}^{[1]}$ in terms of an integral in 
$\lambda$ which involve the flowing coupling $\hat g(\lambda)$
and the rescaling function $Z(\lambda)$
$$
\hat\S_2^{[1]}(0;\Lambda,\Lambda_0)=
Z^{-1}(\Lambda)\int_0^\Lambda\frac{d\lambda}
{\lambda}Z(\lambda)\hat I_2^{[0]}(0;\lambda,\Lambda_0),
$$
\formula{In.riscalate}
{\hat\S_{2n,irr}^{[1]}(p_i;
\Lambda,\Lambda_0)=-Z^{-n}(\Lambda)\int_\Lambda^{\Lambda_0}\frac{d\lambda}
{\lambda}Z^{n}(\lambda)\hat I_{2n,irr}^{[0]}(p_i;\lambda,\Lambda_0).
}
From these vertices one obtains the 
$\hat I_{2n}^{[1]}(p_i;\Lambda,\Lambda_0)$ and constructs the second 
iteration.
After iterating this procedure $\ell$ times, 
$\hat \S_2^{[\ell]}|_{p=0}$ and $\hat\S_{2n,irr}^{[\ell]}$ are given in 
terms of multiple integrals, over the various scales $\lambda_i$ generated 
by the iteration, of complicated expressions involving the flowing coupling 
and the rescaling function at the various scales $\lambda_i$.
In order to obtain the functions $Z(\Lambda)$ and $\hat g(\Lambda)$
needed to compute these integrals, one uses the definitions
$\partial_{p^2}\hat\S_2^{[\ell]}|_{p^2=\mu^2}\equiv1$ and 
$\hat g(\Lambda)=\prendi{\hat\S^{[\ell]}_4}{p_i=\bar p_i}$ which give the 
evolution equations for $Z(\Lambda)$ and $\hat g(\Lambda)$ 
at order $\ell$
\formula{eq.Z}
{\frac{\dot Z}Z=\prendi{\partial_{p^2}\hat I^{[\ell-1]}_2}{p^2=\mu^2},
}
\formula{eq.g}
{\LdL\hat g+2\frac{\dot Z}Z\hat g=
\prendi{\hat I_4^{[\ell-1]}}{p_i=\bar p_i}.
}
These equations are solved with the boundary conditions $Z(0)=1$ and  
$\hat g(0)=g(\mu)$, where $g(\mu)$ is the
coupling constant evaluated at the subtraction point $\mu$.
In general the r.h.s. of \rif{eq.Z} and \rif{eq.g} are functionals
of $\hat g(\lambda_i)$ and $Z(\lambda_i)$ thus one has 
complicated integro-differential equations.

In this  paper we perform the calculation only at the first order, where
\rif{eq.Z} and \rif{eq.g} are simple differential equations, which in 
general can be solved analytically. 
By inserting the solution of these equations in 
\rif{In.riscalate} and setting $\Lambda=0$ one can explicitly 
compute the one-loop improved physical vertices. They are given by the 
same expression of the one-loop perturbative vertices with the 
coupling replaced by the flowing coupling $\hat g(\lambda)$.
This is similar to the standard improved theory in which
one substitutes the coupling with the running coupling $g(q^2)$ in the 
Feynman diagrams. The relation between the two
approaches is discussed in section \ref{other.resum}.

\subsection{Explicit 1-loop calculations}
The rescaling function in the one-loop improved $\phi^4$ theory is trivial
because $\dot Z/Z=\prendi{\partial_{p^2}\hat I_2^{(0)}}\mu\equiv0$,
and therefore $Z(\Lambda)=1$ for any $\Lambda$. 
In the $\Lambda_0\to\infty$ limit
the 1-loop flowing coupling satisfies the evolution equation
\formula{phi4}
{\LdL\hat g(\Lambda)=
\prendi{\hat I_4^{[0]}}{p_i=\bar p_i}
=\frac3{16\pi^2}\hat g^2(\Lambda)F(\Lambda^2/\mu^2),
}
where\footnote
{In the r.h.s. of \rif{phi4} we can take the 
$\Lambda_0\to\infty$ limit because the $\LdL-$derivative 
cuts the higher momenta.}
\formula{def.F}
{F(\Lambda^2/\mu^2)=-16\pi^2\int_q\dot \Delta_{\Lambda\infty}(q)
\Delta_{\Lambda\infty}(q+\bar p),\quad \bar p^2=\mu^2
}
can be exactly calculate specifying the cutoff function.
Notice that $F(\Lambda^2/\mu^2)$ vanishes for $\Lambda=0$ and $F\to 1$ for 
$\Lambda\to\infty$ for any choice of the cutoff function (see appendix A). 
Using the power-law cutoff function given in appendix A (see equation  
\rif{qu.cutoff}) one finds
\formula{FeynDot}
{F(\Lambda^2/\mu^2)=\frac{2\Lambda^2(6\Lambda^4+7\Lambda^2\mu^2+\mu^4)}
{\mu^2(\mu^2+4\Lambda^2)^2}-\frac{48\Lambda^8\mbox{ArcTanh}\sqrt
{\mu^2/(\mu^2+4\Lambda^2)}}{\sqrt{\mu^6}(\mu^2+4\Lambda^2)^{5/2}},
}
with asymptotic limits
$$
F(\Lambda^2/\mu^2)=2\frac{\Lambda^2}{\mu^2}-
2\frac{\Lambda^4}{\mu^4}\quad (\Lambda<<\mu),\quad
F(\Lambda^2/\mu^2)=1-\frac25\frac
{\mu^2}{\Lambda^2}+\frac17\frac{\mu^4}{\Lambda^4}\quad(\Lambda>>\mu).
$$
The solution of equation \rif{phi4} 
\formula{exact.lambda}
{\hat g(\Lambda)=\frac{g(\mu)}{1-\frac3{16\pi^2} g(\mu)
\int_0^\Lambda\frac{d\lambda}
{\lambda} F(\lambda^2/\mu^2)},\quad \hat g(0)= g(\mu)
}
can be expressed in terms of elementary functions but, for sake of 
simplicity, we do not report the lengthy formula. 
In figure 1 we show $\hat g(\Lambda)$ as a function of $\Lambda/\mu$.
\begin{figure}
  \begin{center}
\begin{tabular}{c}   
 \epsfig{file=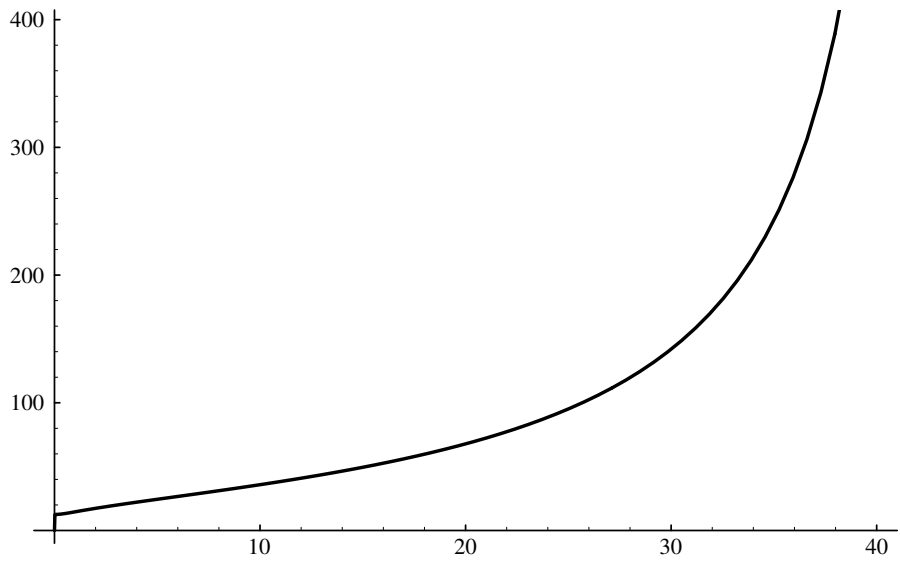,height=90 ex} 
  \end{tabular}
  \end{center}
  \label{1}
  \caption{{\small
Wilsonian flowing coupling $\hat g$ as function of $\Lambda/\mu$
calculated with
the power-law cutoff function. The initial point is $\hat g(0)=
 g=4\pi$, the Landau pole is at $\Lambda=43.47 \mu$}.}
\end{figure}

In a previous article \cite{BMS} we calculated the flowing coupling
with the sharp cutoff function $K_{\Lambda\infty}(q)=\theta(q^2/\Lambda^2-1)$
.~\footnote{With the sharp cutoff we have been able to
evaluate exactly $F(\Lambda^2/\mu^2)$ but not $\hat g(\Lambda)$
.} 
In both cases the flowing coupling has a Landau pole, \ie the denominator
of \rif{exact.lambda} vanishes at finite $\Lambda=\Lambda_L$. 
The existence of the
pole is a general fact because for $\Lambda\geq\bar\Lambda>>\mu$ equation 
\rif{phi4} has the {\it universal} (\ie independent on the cutoff function) 
asymptotic solution 
$$
\hat g_{as}(\Lambda)=
\frac{\hat g(\bar\Lambda)}{1-\frac3{16\pi^2}\hat g(\bar\Lambda)\log\Lambda/
\bar\Lambda}.
$$
However, the exact position of the Landau pole
depends on the value of $g(\mu)$ and on
the choice of the cutoff function. For instance, by fixing $g(\mu)=4\pi$,
with the power-law cutoff one has $\Lambda_L=43.47\mu$ while with the 
sharp cutoff one has $\Lambda_L=39.79\mu$.
We have also  computed the pole position in the case of the 
exponential cutoff $K_{\Lambda\infty}(q)=1-\exp(-q^2/\Lambda^2)$
obtaining $\Lambda_L=37.74\mu$.

By inserting in \rif{In.riscalate} 
the flowing coupling calculated from \rif{exact.lambda} and
taking $\Lambda=0$ one obtains the physical one-loop 
improved vertices. Notice that the presence of the Landau pole at 
$\Lambda=\Lambda_L$
implies that the ultraviolet cutoff $\Lambda_0$ cannot be removed in this
case, although the theory is perturbatively renormalizable. This corresponds
to the property of triviality, entailing that the $\Lambda_0\to\infty$ limit
is possible only if the coupling $g(\mu)$ vanishes.

\section{\bom{\phi^3_6} theory}
The rescaling function $Z(\Lambda)$
was not involved in the previous discussion on
the $\phi^4_4$ theory at one loop. Therefore, before treating QED, it is
useful to 
consider the (massive) $\phi^3_6$ theory in six dimension as an example 
in which $Z(\Lambda)$ is not trivial at one loop. This theory is interesting 
even because it mimics some features of QCD (\ie it is an asymptotically 
free theory).
Moreover, we analyze the presence of a fixed mass $m$ different from zero.

The improved tree level cutoff action is
$$
\hat\Gamma^{(0)}=\frac12 \hat\phi\cdot\PropI\hat\phi+\frac1{3!}
\int_x \hat g(\Lambda)\hat\phi^3,\quad
\Prop(q)\equiv\frac{K_{\Lambda\Lambda_0}(q)}{q^2+m^2}
$$
and the one-loop improved evolution equations in the $\Lambda_0\to\infty$ 
limit read
$$
\LdL\hat\S_2+\frac{\dot Z}Z\hat\S_2=-
\hat g^2(\Lambda)
\int_q\dot \Delta_{\Lambda\infty}(q) \Delta_{\Lambda\infty}(q+p),
$$
$$
\LdL\hat\S_3+\frac32\frac{\dot Z}Z\hat\S_3=3\hat g^3(\Lambda)
\int_q\dot\Delta_{\Lambda\infty}(q) \Delta_{\Lambda\infty}(q+p) \Delta_
{\Lambda\infty}(q+p+p')
$$
and similarly for other vertices\footnote{We do not consider the $n=1$ vertex 
because it is momentum independent and then vanishes by zero-momentum 
subtraction.}.
From the definitions $\prendi{\partial_{p^2}\hat\S_2}{p=0}
\equiv1$ and $\hat g\equiv\prendi{\hat\S_3}{p_i=0}$ one has
\formula{diff.eq.phi3}
{\frac{\dot Z}Z=-\frac16r\hat g^2 F_\phi,\quad \LdL\hat g=-r \hat g^3 F_g-
\frac32\frac{\dot Z}Z\hat g,\quad r=\frac1{(4\pi)^3}
}
where
\formulona{F.6d}
{F_\phi(\Lambda)=6(4\pi)^3\prendi{\partial_{p^2}\int_q
\dot\Delta_{\Lambda\infty}(q)\Delta_{\Lambda\infty}(q+p)}{p=0},\\
F_g(\Lambda)=-3(4\pi)^3\int_q\dot\Delta_{\Lambda\infty}(q)\Delta_{\Lambda
\infty}(q)^2
} 
are functions growing from 0 to 1 which can be exactly calculated 
specifying the cutoff function.
From \rif{diff.eq.phi3} one finds that the flowing coupling 
$\hat g$ is determined by the differential equation
\begin{figure}
  \begin{center}
    \begin{tabular}{c}
    \epsfig{file=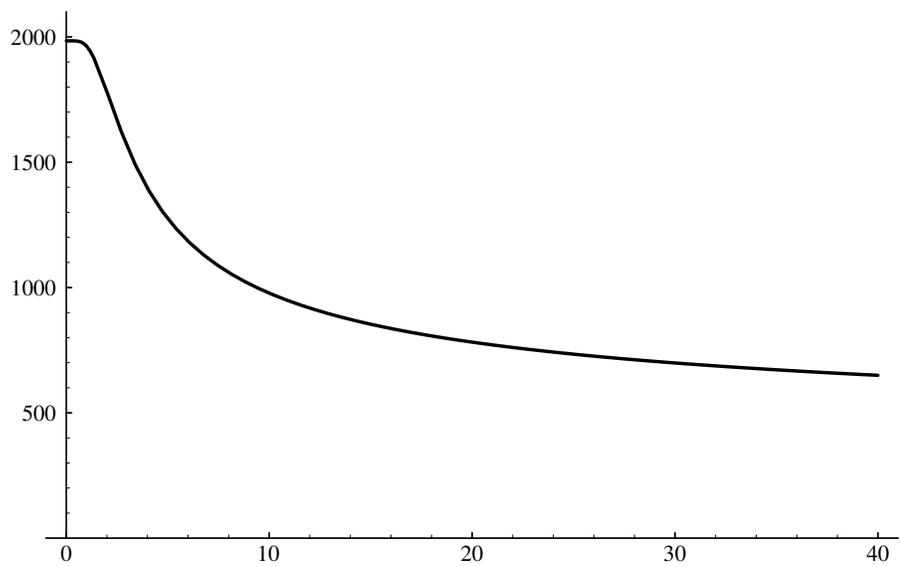,height=90 ex} 
    \end{tabular}
  \end{center}
  \label{2}
  \caption{{\small
Wilsonian flowing coupling $\hat g^2$ as function of $\Lambda/m$
in $\phi^3_6$ theory
calculated using the power-law cutoff function. The initial point is 
$\hat g^2(0)=g^2=(4\pi)^3$}.}
\end{figure}
\formula{ex.eq.phi3}
{\LdL\hat g=[-F_g(\Lambda)+\frac14F_\phi(\Lambda)]r\hat g^3,\quad \quad 
\hat g(0)=g.
}
We solved this equation using the functions 
$F_\phi(\Lambda)$ and $F_g(\Lambda)$ 
computed with the power-law cutoff (see appendix A) and 
the result is displayed in figure 2.
In the ultraviolet limit $\Lambda\geq\bar\Lambda>>m$ equation  
\rif{ex.eq.phi3} becomes 
$$
\LdL\hat g_{as}=-\frac34r\hat g_{as}^3=-b_1\hat g_{as}^3,\quad \quad 
b_1=\frac3{256\pi^3}
$$
and therefore the asymptotic flowing coupling has the same functional form of
the running coupling 
$$
g_{as}^2(\Lambda)=\frac{\bar g^2}{1+b_1\bar g^2\log\Lambda/\bar\Lambda},
\quad \quad g_{as}(\bar\Lambda)=\bar g.
$$
Notice that the asymptotic coupling is ill defined in the infrared for
$\Lambda=\Lambda_L=\bar\Lambda e^{-1/(b_1 \bar g^2)}$ (Landau pole), 
on the contrary the flowing $\hat g(\Lambda)$ is regular for any $\Lambda$.
From \rif{diff.eq.phi3} one also obtains the rescaling function
$$
Z(\Lambda)=\exp[-\frac16r \int_0^\Lambda\frac{d\lambda}{\lambda}\hat g^2
(\lambda)F_\phi(\lambda)]
$$
which is well defined for any $\Lambda$ and decreases to
zero for $\Lambda\to\infty$. In the asymptotic limit one has
$Z(\Lambda)\to \left(1+b_1 \bar g^2\log\Lambda/\bar\Lambda\right)^
{-\frac r{6 b_1}}Z(\bar\Lambda)$.

\section{Improved QED}
In this section we generalize the previous calculations to the QED case. 
The crucial point is to show that the improved
theory at $\Lambda=0$ satisfies the Ward identities associated with the 
gauge transformation
$$
\delta_\ep\psi(x)= i e \ep(x)\psi(x),\quad \delta_\ep\bar\psi(x)=-
ie\bar \psi(x)\ep(x),\quad
\delta_\ep A_\mu(x)=\partial_\mu \ep(x).
$$
The tree level improved cutoff action is
\formula{QED.impr}
{\hat\Gamma^{(0)}(\hat\phi,\hat a,\hat e;\Lambda,\Lambda_0)
=\int_x\frac12 \hat A_\mu (D^{-1}_{\Lambda\Lambda_0})^{\mu\nu}
\hat A_\nu+\hat{\bar\psi}S^{-1}_{\Lambda\Lambda_0}\hat \psi+\hat e(\Lambda) 
\hat{\bar\psi}\hat{\slash A}\hat \psi,
}
where $\hat e(\Lambda)$ is the flowing coupling and we have introduced 
the rescaled fields $\hat\psi=Z^{1/2}_\psi \psi$, 
$\hat{\bar\psi}=Z^{1/2}_\psi \bar\psi$ 
and $A_\mu=Z_A^{1/2}A_\mu$. The cutoff propagators are
$$
D_{\Lambda\Lambda_0,\mu\nu}(k)=-
\left(\frac{g_{\mu\nu}}{k^2}-(1-\hat a(\Lambda))\frac{k_\mu k_\nu}
{k^4}\right)K_{\Lambda\Lambda_0}(k)
$$
and
$$
S_{\Lambda\Lambda_0}(p)=-
\frac{\slash p+m}{p^2-m^2} K_{\Lambda\Lambda_0}(p),
$$
where $K_{\Lambda\Lambda_0}(k)$ and $K_{\Lambda\Lambda_0}(p)$,
after  analytic continuation in euclidean space, become the 
power-law cutoff functions defined in appendix A. 
In particular for the photon propagator we use the cutoff function 
\rif{qu.cutoff} while for the electron propagator we use the massive 
cutoff function \rif{qu.cutoff'}. 
Notice that the photon propagator contains the $\Lambda-$dependent
gauge fixing parameter $\hat a(\Lambda)$. As we will see, this is required by
gauge invariance.

The evolution equations for the rescaled vertices are obtained 
following the same steps discussed in Section 3. In this case
the measure $\hat M$ (corresponding to \rif{misura}) contains both the
electron and photon propagators.  
In the following we use the notation 
$\LdL D_{\Lambda\Lambda_0,\mu\nu}$ for the contributions to $\hat M$ 
coming from the photon propagator even though 
the derivative acts only on the cutoff function and not on 
the gauge fixing parameter $\hat a(\Lambda)$. 

Starting from \rif{QED.impr} one 
can iteratively construct the improved
vertex functions at higher orders.
The rescaling functions are obtained
solving the corresponding evolution equations with boundary conditions 
$Z_A(0)=1,\ Z_\psi(0)=1$, while $\hat e(\Lambda)$ and $\hat a(\Lambda)$
are not independent functions. Indeed gauge invariance requires 
the flowing coupling $\hat e(\Lambda)$ to be related to the rescaling 
function $Z_A(\Lambda)$ by
\formula{hat.e}
{\hat e(\Lambda)= e Z_A^{-1/2}(\Lambda).
} 
Similarly the flowing
gauge fixing coupling $\hat a(\Lambda)$ must satisfy the relation 
$\hat a(\Lambda)= a Z_A(\Lambda)$, where $a$ is a fixed number  which does not
affect the physics (for instance $a=1$ in the Feynman gauge). 
In this way the tree level improved action in terms of the non-rescaled fields
$$
\S^{(0)}=\int_x-Z_A(\Lambda)[\frac14F_{\mu\nu} F^{\mu\nu}+
\frac1{2\hat a(\Lambda)}(\partial\cdot A)^2]+
Z_\psi(\Lambda)\bar\psi(i\slash\partial+ \hat e Z_A^{1/2}\slash{A}-m)\psi
$$
satisfies the standard Ward identity
$$
W(x)\S^{(0)}=(-\partial_\mu\dd{\S^{(0)}}{A_\mu}-
i e\bar\psi\dd{\S^{(0)}}{\bar\psi}+i e\dd{\S^{(0)}}\psi\psi)=
\frac{\Box}{a}\partial \cdot A(x)
$$
at any scale $\Lambda$.
Through this property we will prove that the physical
improved action $\S^{[1]}(\phi;0,\Lambda_0)$
satisfies Ward identities up order  $\O(1/\Lambda_0)$, then in the limit
$\Lambda_0\to\infty$ 
``gauge-invariance'' is preserved by the improved perturbative expansion, at
least at one loop.

\vskip 10pt
\noindent
We now perform some explicit computations.
We denote by $\hat\S^{[1]}_{\mu_1\dots\mu_n\alpha_1\dots\alpha_{2m}}
(p_i;\Lambda,\Lambda_0)$ the one-loop improved vertices 
with $n$ photons and $m$ pairs of fermions.

The (inverse) photon propagator evolution equation is
\formula{photon.eq}
{(\LdL+\frac{\dot Z_A}{Z_A})\hat\S_{\mu\nu}^{[1]}(p;\Lambda,\Lambda_0)=
\hat e^2(\Lambda)\hat I_{\mu\nu}(p;\Lambda,\Lambda_0),
}
where
$$
\hat I_{\mu\nu}(p;\Lambda,\Lambda_0)=i\int_q\LdL\Tr(\gamma_\mu S_{
\Lambda\Lambda_0}(q)\gamma_\nu S_{\Lambda\Lambda_0}(q+p))
$$
and we have explicitly written the dependence on $\hat e(\Lambda)$.
From this equation and the normalization condition 
$-\prendi{\partial_{p^2}\frac13t^{\mu\nu}\hat\S_{\mu\nu}}{p^2=0}=1$,
where $t^{\mu\nu}= g^{\mu\nu}-p^\mu p^\nu/p^2$, 
one obtains the evolution equation for the rescaling function 
$Z_A(\Lambda)$. In particular choosing the power-law cutoff function 
and taking the limit $\Lambda_0\to\infty$ one has 
$$
\frac{\dot Z_A}{Z_A}=-\hat e^2(\Lambda)
\prendi{\partial_{p^2}\frac13t^{\mu\nu}\hat I_{\mu\nu}}{p^2=0}=
- \frac{\hat e^2(\Lambda)}{6\pi^2}
\frac{\Lambda^4(5\Lambda^4+14\Lambda^2 m^2+6 m^4)}{5(\Lambda^2+m^2)^4}.
$$
Using \rif{hat.e} and the boundary condition $Z(0)=1$ one finds 
$$
Z_A(\Lambda)=1-\frac{e^2}{6\pi^2}\left[\frac12\log
\frac{\Lambda^2+m^2}{m^2}-\frac{\Lambda^2(7\Lambda^4+19\Lambda^2 m^2+10 m^4)}
{20(\Lambda^2+m^2)^3}\right].
$$
In figures 3 and 4 we display the flowing couplings 
$\hat \alpha(\Lambda)=\frac{\hat e^2(\Lambda)}{4\pi}$ and 
$\hat a(\Lambda)= a Z_A(\Lambda)$ 
calculated with this choice of the cutoff function.
Notice that the coupling $\hat\alpha(\Lambda)$ goes to
infinity for a finite value $\Lambda=\Lambda_L$, the Landau pole. 
The exact position of the pole depends on the
choice of $K_{\Lambda\infty}(q)$. 
\begin{figure}
  \begin{center}
  \begin{tabular}{c}
    \epsfig{file=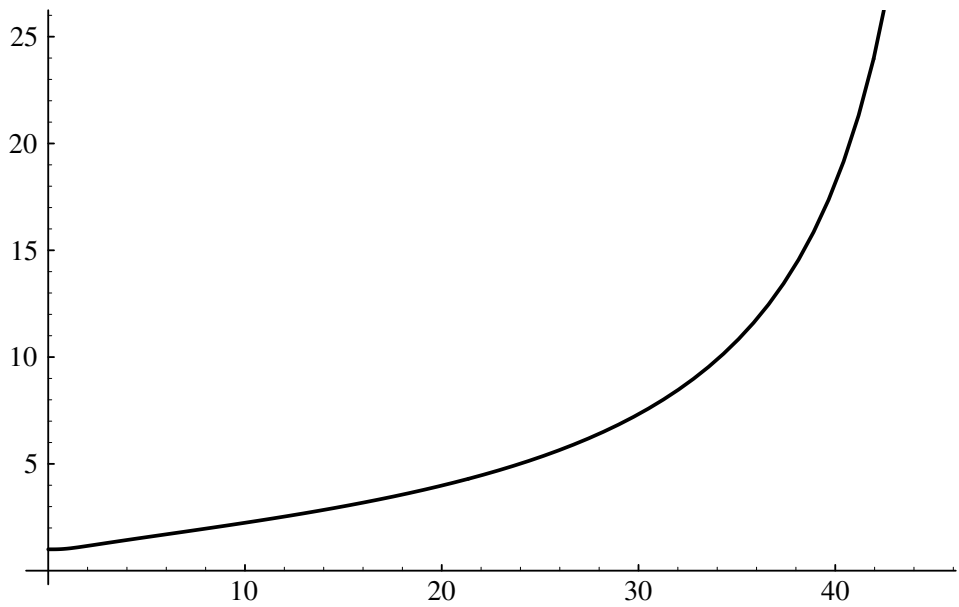,height=96 ex}\\
  \end{tabular}
  \end{center}
  \label{3}
  \caption{{\small
   Wilsonian flowing coupling $\hat \alpha$ as function of $\Lambda/m$
  in QED calculated using the power-law cutoff function. 
  The initial point is $\hat \alpha(0)=\alpha=1$.}}
\end{figure}
\begin{figure}
  \begin{center}
  \begin{tabular}{c}
      \epsfig{file=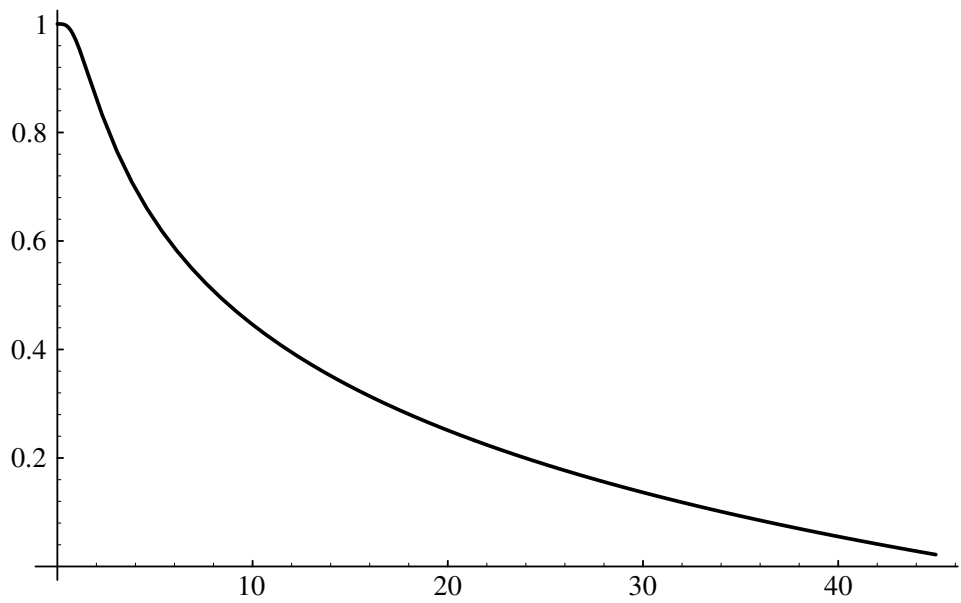,height=96 ex}
    \end{tabular}
  \end{center}
  \label{4}
  \caption{{\small
  Wilsonian flowing coupling $\hat a$ as function of $\Lambda/m$. 
  The initial point is $\hat a(0)=a=1$ (Feynman gauge).}}
\end{figure}

The electron rescaling function can be obtained from the 
(inverse) electron propagator evolution equation
\formula{electron.eq}
{(\LdL+\frac{\dot Z_\psi}{Z_\psi})\hat\S_{\alpha\beta}^{[1]}(p;\Lambda,
\Lambda_0)=\hat e^2(\Lambda)\hat I_{\alpha\beta}(p;\Lambda,\Lambda_0),
}
where 
$$
\hat I_{\alpha\beta}(p;\Lambda,\Lambda_0)
=-i\int_q\LdL\left[\gamma_\mu S_{\Lambda\Lambda_0}(q+p)
\gamma_\nu D^{\mu\nu}_{\Lambda\Lambda_0}(q)\right]_{\alpha\beta},
$$
and the normalization condition $\prendi{\partial_{p^\mu}
\hat\S_{\alpha\beta}}{p^2=0}=\gamma^\mu_{\alpha\beta}$. Also in
this case, using the power-law cutoff function, one can compute 
explicitly $Z_\psi(\Lambda)$.
Here we give only its asymptotic limit for $\Lambda_L>>\Lambda>>m$ 
$$
Z_\psi(\Lambda)\simeq 1-\frac1{16\pi^2} a e^2\log\frac{\Lambda^2}{m^2}. 
$$
For further references we report the vertex evolution equation
\formula{vertex.eq}
{(\LdL+\frac{\dot Z_\psi}{Z_\psi}+\frac12\frac{\dot Z_A}{Z_A})
\S_{\mu\alpha\beta}^{[1]}(p,p';\Lambda,\Lambda_0)
=\hat e^3(\Lambda) I_{\mu\alpha\beta}(p,p';\Lambda,\Lambda_0),
}
where
$$
I_{\mu\alpha\beta}(p,p';\Lambda,\Lambda_0)=-i
\int_q\LdL\left[\gamma^\rho S_{\Lambda\Lambda_0}(q+p)
\gamma_\mu S_{\Lambda\Lambda_0}(q+p')\gamma^\sigma 
D_{\Lambda\Lambda_0,\sigma\rho}(q)\right]_{\alpha\beta}.
$$

\section{Gauge invariance of the improved theory} \label{gauge-inv}
In this section, we  explicitly check the gauge invariance of the one-loop
improved vertices  at $\Lambda=0$. 
In particular we analyze the transversality of 
the photon propagator and the Ward identity for the vertex.  
The longitudinal part of the photon propagator   
$\S_L\equiv\frac{p^\mu p^\nu}{p^2}\S_{\mu\nu}$ is obtained from 
\rif{photon.eq} and it is given by
\formula{long.part}
{\S_L^{[1]}(p;0,\Lambda_0)=
-\int_0^{\Lambda_0}\frac{d\lambda}{\lambda}Z_A(\lambda)\hat 
e^2(\lambda)\hat I_{L,irr}(p;\lambda),}
where
\formulaX{\hat I_{L,irr}(p;\lambda)=\hat I_{L}
-\prendi{\hat I_{T}}{p=0}-p^2\prendi{\partial_{\bar p^2}
\hat I_L}{\bar p^2=0}, \quad\hat I_{L}=\frac{p^\mu p^\nu}{p^2}\hat I_{\mu\nu},
\quad\hat I_{T}=\frac13t^{\mu\nu}\hat I_{\mu\nu}
}
(here and in the following the $\Lambda_0-$dependence in $\hat I_L$ is 
understood).
The subtractions in the integrand of \rif{long.part} are a consequence 
of isolating the relevant couplings in $\S_{\mu\nu}$ and of the 
different boundary conditions (see appendix B for details).
By using \rif{hat.e} one has that the $Z_A(\lambda)$ factors 
in \rif{long.part} cancel and the only dependence on $\lambda$ in the
integral is in the cutoff propagators. Therefore $\S_L^{[1]}$ 
is equal to the longitudinal part of the photon
propagator obtained in the standard perturbation theory,
which for large $\Lambda_0$ vanishes as negative powers of $\Lambda_0$ 
\cite{qed}. 
In perturbation theory renormalizability ensures that 
$\Lambda_0$ can be sent to infinity and then the longitudinal part of 
the photon propagator vanishes. 
In the improved perturbation theory the presence on the Landau pole 
at $\Lambda_L$ implies that $\Lambda_0$ cannot be removed. 
Therefore in this case one recovers the transversality of the 
photon propagator only for momenta much lower than $\Lambda_0$.

The violation of the Ward identity for the vertex 
is given by the following quantity 
\formula{W3}
{\Delta_{\alpha\beta}^{[1]}(p,p';0,\Lambda_0)=(p'-p)^\mu\S_{\mu
\alpha\beta}^{[1]}(p,p';0,\Lambda_0)-e\S^{[1]}_{\alpha\beta}(p';0,\Lambda_0)+
e\S^{[1]}_{\alpha\beta}(p;0,\Lambda_0).
}
Using \rif{electron.eq}, \rif{vertex.eq} and \rif{hat.e} one obtains
\formula{W33}
{\Delta_{\alpha\beta}^{[1]}
=e^3\int_0^{\Lambda_0}\frac{d\lambda}{\lambda}\frac{Z_\psi(\lambda)}
{Z_A(\lambda)}[
(p'-p)^\mu\hat I_{\mu\alpha\beta,irr}(p,p';\lambda)
-\hat I_{\alpha\beta,irr}(p';\lambda)
+\hat I_{\alpha\beta,irr}(p;\lambda)],
}
where 
$$
\hat I_{\mu\alpha\beta,irr}(p,p';\lambda)=
\hat I_{\mu\alpha\beta}(p,p';\lambda)-\hat 
I_{\mu\alpha\beta}(0,0;\lambda)
$$
and
$$
\hat I_{\alpha\beta,irr}(p;\lambda)=\hat I_{\alpha\beta}(p;\lambda)
-\hat I_{\alpha\beta}(0;\lambda)-p^\mu\partial_{\bar p^\mu}\hat I_{\alpha\beta}
(0;\lambda).
$$
Notice that setting $Z_A=Z_\psi=\hat a=1$ for any $\lambda$, equation 
\rif{W33} becomes the violation of the usual 
perturbation theory, which vanishes for $\Lambda_0\to\infty$ as shown 
in ref.~\cite{qed}. The proof is based on the following 
identity 
$$
\frac1{\slash q+\slash p+m}(\slash p'-\slash p)\frac1{\slash q+\slash p'+m}=
\frac1{\slash q+\slash p+m}-\frac1{\slash q+\slash p'+m}
$$
and on the fact that in this case the integrand is a total derivative 
(see \rif{electron.eq} and \rif{vertex.eq}) so that the result of the 
integration over $\lambda$ is 
\formulona{viol0}
{-i e^3 \int_q\frac{K_{0\Lambda_0}(q)}{q^2}\gamma_\rho&\!\!\!\!\Bigl\{
\frac{K_{0\Lambda_0}(q+p)}
{\slash q+\slash p+m}\gamma^\rho\left[K_{0\Lambda_0}(q+p')-1\right]\\
&+\frac1{\slash q+m}\gamma_\rho p_\mu\left[\dede{}{\bar p_\mu}K_{0\Lambda_0}
(q+\bar p)\right]_{\bar p=0}\Bigr\}-(p\to p').
}
For $\Lambda_0\to\infty$ this integral vanishes as negative powers 
of $\Lambda_0$. 

In the improved theory the vanishing of 
$\Delta_{\alpha\beta}^{[1]}$ can be proved in a similar way. 
Consider first the contribution in \rif{W33} coming from 
the $g_{\mu\nu}$ part of the photon propagator.
One can apply the mean value theorem to extract from the integral the factor 
$Z_\psi(\bar\lambda)/Z_A(\bar\lambda)$ for some scale $\bar\lambda$, 
with $0\leq\bar\lambda\leq\Lambda_0$. 
This factor multiplies the same integral of the non-improved case.
Therefore for large $\Lambda_0$ (but  $\Lambda_0<<\Lambda_L$) 
this contribution vanishes independently 
of $\bar\lambda$ since $Z_A(\bar\lambda)$ and  $Z_\psi(\bar\lambda)$ 
are at most logarithmically divergent while \rif{viol0} vanishes 
as negative powers of $\Lambda_0$.
The remaining contribution can be treated in the same way. 
By applying the mean value theorem one extracts the factor 
$\frac {Z_\psi}{Z_A}(\hat a-1)$ at some scale $\bar\lambda$, so that 
also in this case the integrand is a total derivative 
and the result of the $\lambda$ integration is 
\formulona{viol'}
{-i e^3\int_q\frac{K_{0\Lambda_0}(q)}{q^4}\slash q&\!\!\!\!\Bigl\{
\frac{K_{0\Lambda_0}(q+p)}
{\slash q+\slash p+m}\slash q\left[K_{0\Lambda_0}(q+p')-1\right]\\
&+\frac1{\slash q+m}\slash q p_\mu\left[\dede{}{\bar p_\mu}K_{0\Lambda_0}
(q+\bar p)\right]_{\bar p=0}\Bigr\}-(p\to p').
}
For the argument given above also in this case the result
of the $q-$integration vanishes as negative power of $\Lambda_0$.
As noted above one cannot remove $\Lambda_0$ due to the Landau pole and 
therefore also the vertex Ward identity is valid 
only in a weak sense \ie for momenta much lower than $\Lambda_0$.

\section{Comparison with the standard improved formulation}\label{other.resum}
In this section we compare our improved perturbation theory 
with the standard formulation. We show that for non-asymptotically
free theories our formulation is very similar to the standard one.
To this aim we consider a simple example which trivially generalizes to
other interesting cases, \ie the calculation of the improved fish diagram.
In the standard resummation
approach \cite{renorm} one passes from a one-loop perturbative
quantity to an improved quantity simply replacing the coupling $g$ 
in the vertices of the Feynman diagrams with the one-loop running
coupling $g(q^2)$, where $q$ is the momentum flowing in the loop.
This modification means that one has to consider quantities such as 
\formula{standard.impr}
{\tilde\F(Q^2/\mu^2;\Lambda_0^2/\mu^2)=\int_0^{\Lambda_0^2}
\frac{dq^2}{2 q^2} g^2(q^2)[\tilde F(q^2/Q^2)-\tilde F(q^2/\mu^2)]
}with
\formula{F.l}
{\tilde F(q^2/Q^2)=
\frac{q^4}{2\pi^2}\int_\Omega\frac1{q^2}\frac1{(q+Q)^2},
}
where $\int_\Omega$ indicates the angular integral in four dimension
and the $q^4$ factor gives us a dimensionless quantity.
On the other hand, in our formulation one has to calculate integrals of
the form
\formula{our.impr}
{\F(Q^2/\mu^2;\Lambda_0^2/\mu^2)=\int_0^{\Lambda_0^2}
\frac{d\lambda^2}{2\lambda^2} \hat g^2(\lambda)[F(\lambda^2/Q^2)-
F(\lambda^2/\mu^2)]
} 
where $F$ is the function given by \rif{def.F}. 
We want to show that \rif{standard.impr} and \rif{our.impr} are 
numerically almost the same, \ie $F\simeq\tilde F$. In this simple example
\rif{F.l} can be exactly calculated and one gets
$$
\tilde F(\Lambda^2/Q^2)=\frac{2\Lambda^2}
{Q^2+4\Lambda^2}+\frac{8\Lambda^4\mbox{ArcTanh}
\sqrt{Q^2/(Q^2+4\Lambda^2)}}{\sqrt{Q^2} (Q^2+4\Lambda^2)^{3/2}}.
$$
Comparing this result with \rif{FeynDot} one finds 
that $F$ and $\tilde F$ have the same asymptotic limits
$\Lambda\to0$ and $\Lambda\to\infty$
$$
|\tilde F- F|\to\frac{6\Lambda^4}{Q^4}\to0, \quad \Lambda^2<<Q^2,
$$
$$
|\tilde F- F|\to \frac1{15}\frac{Q^2}{\Lambda^2}\to0, \quad \Lambda^2>>Q^2
$$
and that the relative error
$\ep(\Lambda^2/Q^2)=|\tilde F-F|/|\tilde F|$ is small in any momentum range.
The plot of $F$ and $\tilde F$ is reported in figure 5.

\begin{figure}
  \begin{center}
  \begin{tabular}{c}
    \epsfig{file=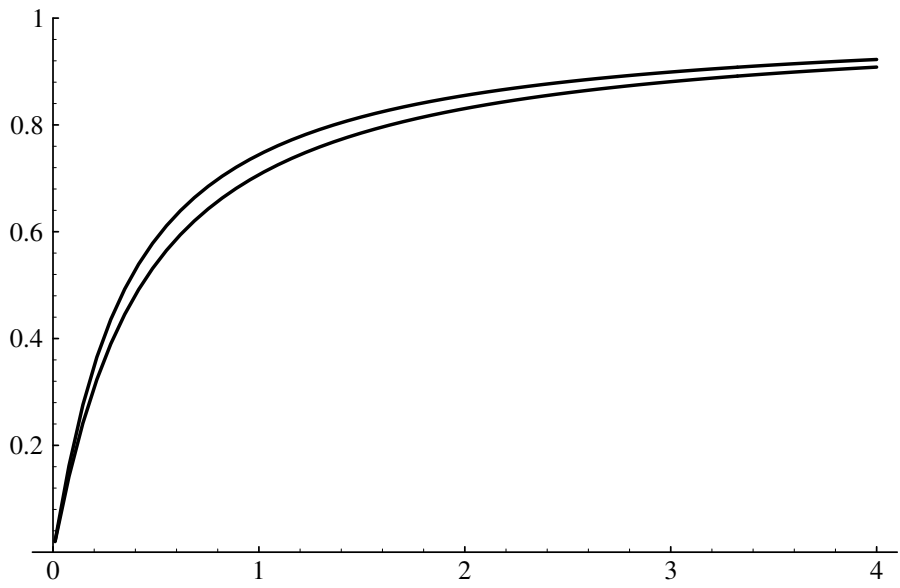,height=90 ex} 
    \end{tabular}
  \end{center}
  \label{5}
  \caption{{\small Comparison between the exact result $F(\Lambda^2/Q^2)$ 
  (top curve) and the leading result $\tilde F(\Lambda^2/Q^2)$ 
  (bottom curve)}.}
\end{figure}
The fact that the function $F$ is not so different from $\tilde F$ 
can be also seen using in \rif{def.F} the sharp cutoff function 
$K_{\Lambda\infty}(q)=\theta(q^2/\Lambda^2-1)$ and the approximation
$K_{\Lambda\infty}(q+Q)\simeq K_{\Lambda\infty}(q)$. 
In this case one gets
$$
F(\Lambda^2/Q^2)\simeq-8\pi^2
\int_q\LdL\frac{\theta(q^2/\Lambda^2-1)}{q^2(q+Q)^2}=\tilde
F(\Lambda^2/Q^2),
$$
where we have used the properties 
$K_{\Lambda\infty}(q)^2=K_{\Lambda\infty}(q)$ and 
$\dot K_{\Lambda\infty}(q)= -2\Lambda^2\delta(q^2-\Lambda^2)$.
This argument generalizes to one-loop graphs with an arbitrary number 
of cutoff propagators and various indices (Lorentz, spinor, color, etc) 
even though these integrals in general cannot be explicitely calculated.

In the approximation $F\simeq \tilde F$
our improved theory is completely equivalent to the standard improved theory,
with the only difference of replacing the running coupling $g(q^2)$
with the flowing coupling $\hat g(\lambda)$.~\footnote
{In QED the analogous
of \rif{our.impr} also contains the electron rescaling function, 
but $Z_\psi(\lambda)\simeq1$ in the $\lambda<<\Lambda_L$ region. }
Notice that in the ultraviolet region $g(\sqrt{q^2})$ and $\hat g(\lambda)$ 
have the same functional form, while in the infrared they differ. 
In particular in the case of asymptotically free theories 
$\hat g(\lambda)$ is regular in all the range of $\lambda$ and 
the integral \rif{our.impr} is well defined while \rif{standard.impr} 
suffers for the infrared Landau pole. 

\section{Conclusion}
We have formulated a systematic improved perturbation theory, based on 
the Wilson renormalization group. In this approach one solves 
iteratively the $\Lambda-$RG equations in terms of the Wilsonian 
flowing coupling $\hat g(\lambda)$. 
In the ultraviolet region this coupling becomes the running coupling 
constant and our formulation becomes equivalent to 
the standard improved perturbation theory \cite{renormalons}.

In this paper we have considered the $\phi^3_6$ theory and we have 
shown that the flowing coupling is finite in all the range of $\lambda$ so 
that our improved perturbation theory is well defined. 
On the contrary the standard improved perturbation theory 
is ambiguous due to the infrared Landau pole. 
We consider this simple model but this result must hold also for 
the Yang-Mills and QCD theories.
A preliminary study of a non-Abelian gauge theory indicates that this 
is indeed the case \cite{BMS}.

The analysis of a gauge theory requires some care 
due to the issue of gauge invariance.  
In this paper we have considered the QED case and we 
have proved that our improved formulation is consistent with Ward 
identities, although there are breaking terms because 
the ultraviolet cutoff cannot be removed. 
This is due to the presence of the ultraviolet Landau 
pole in the flowing coupling which reflects the 
effective character of this theory. 
Therefore we were able to prove that the first iteration gives 
a photon two-point function which is transverse only for momenta 
much lower than the Landau pole. 
Similarly  the Ward identity for the photon-electron 
vertex is satisfied in this limit.
The essential ingredient of the proof is the gauge invariance of the 
starting point of the iteration. 
As a conseguence the flowing couplings $\hat e(\lambda)$ and $\hat 
a(\lambda)$ are related to the rescaling function $Z_A(\lambda)$.
Once the tree level gauge invariance is implemented,
the proof follows the same steps of the 
perturbative case and therefore can be extended to all vertices 
and, we believe, to all iterations. 

The most interesting case is the non-Abelian gauge theory, in which 
there are several indications that the running coupling constant 
becomes the effective coupling to use in the Feynman diagrams, but no 
systematic proof of this assumption is known \cite{BB}.
The generalization of our method to the non-Abelian case and the proof 
of the Slavnov-Taylor identities for the one-loop improved effective action 
is under study. 

\vspace{1cm}\noindent{\large\bf Acknowledgments}
\vskip 0.3 true cm

We would like to thank F. Vian for discussions and 
G. Marchesini for many suggestions and careful reading part of the manuscript.

\vskip 1 true cm
\noindent
{\Large\bf Appendix A}
\vskip 0.3 true cm
\noindent
In this appendix we give a method to calculate general cutoff 
Feynman graphs, specifying an useful form for the cutoff function.
The simplest integral one has to compute is
\formula{integr.tipico}
{F(\Lambda^2/p^2)=-8\pi^2\int_q\LdL\left[\frac{K_{\Lambda\infty}(q)}{q^2}
\frac{K_{\Lambda\infty}(q+p)}{(q+p)^2}\right] .
}
This integral can be exactly calculated in polar coordinates
using the sharp cutoff function (see \cite{Morris,BMS})
\formula{sharp.cutoff}
{K_{\Lambda}(q)=\theta(q^2/\Lambda^2-1).
}
However, with this cutoff function one cannot compute the integrals with 
more propagators for any $\Lambda$ and for general configurations
of the momenta.
Moreover the sharp cutoff function \rif{sharp.cutoff}
is not differentiable and requires  some care \cite{Morris}. 
To avoid all these problems in this paper
we consider the power-law cutoff function
\formula{qu.cutoff}
{K_{\Lambda\infty}(q)=1-K_{0\Lambda}(q),\quad K_{0\Lambda}(q)=\frac{\Lambda^4}
{(q^2+\Lambda^2)^2}.
}
The cutoff function with both ultraviolet and infrared cutoffs is 
given by $K_{\Lambda\Lambda_0}\equiv K_{0\Lambda_0}-K_{0\Lambda}$.
With this choice of the cutoff function the integral 
\rif{integr.tipico} becomes
$$
F(\Lambda^2/p^2)=64\pi^2
\int_q\frac{\Lambda^4}{(q^2+\Lambda^2)^3}
\frac{(q+p)^2+2\Lambda^2}{((q+p)^2+\Lambda^2)^2}
$$
which can be computed using Feynman parameterization
formulae. The generalization to integrals with more propagators is 
straightforward. 

Notice that $F(\L^2/p^2)\to 1$ for large $\Lambda>>p$.
This behaviour is universal, \ie independent of the precise form 
of the cutoff function. Indeed setting 
$K_{\Lambda\infty}(q)\equiv k(x)$, with $x=q^2/\Lambda^2$, 
one has
$$
-8\pi^2\int_q\LdL \left(\frac{K_{\Lambda\infty}(q)}{q^2}\right)^2=
-\frac12\int_0^\infty \frac{dx}x[-2x\partial_x (k(x)^2)]=1
$$
since any cutoff function satisfies to $k(0)=0$ and $k(\infty)=1$. 
For massive fields it is convenient to use the mass-dependent cutoff 
function
\formula{qu.cutoff'}
{K_{\Lambda\infty}(q)=1-\frac{\Lambda^4}{(q^2+m^2+\Lambda^2)^2}
}
since in this way the Feynman integrals can be calculated using the 
Feynman parametrization formulae as in the massless case.

The power-law cutoff function \rif{qu.cutoff'} can be extended 
to $d$ dimensions, $d>2$, defining
$$
K_{\Lambda\infty}(q)=1-\left(\frac{\Lambda^2}{q^2+m^2+\Lambda^2}
\right)^{[d/2]},
$$
where $[d/2]$ indicates the integer part of $d/2$. 
In particular in $d=6$ one has 
\formula{cutoff.6d}
{K_{\Lambda\infty}(q)=
(q^2+m^2)\frac{(q^2+m^2)^2+3\Lambda^2(q^2+m^2+\Lambda^2)}
{(q^2+m^2+\Lambda^2)^3},
}
which has been used in section~4 to compute the flowing couplings of the 
$\phi_6^3$ theory. In particular the functions $F_\phi$ 
and $F_g$ defined in \rif{F.6d} and computed using 
\rif{cutoff.6d} are given by
$$
F_\phi(\Lambda)=\frac{\Lambda^6(35\Lambda^4+40\Lambda^2m^2+14m^4)}
{35(\Lambda^2+m^2)^5}
$$
$$
F_g(\Lambda)=
\frac{\Lambda^6(140\Lambda^8+381\Lambda^6m^2+414\Lambda^4m^4+210
\Lambda^2m^6+42m^8)}{140(\Lambda^2+m^2)^7}. 
$$

\vskip 1 true cm
\noindent
{\Large\bf Appendix B}
\vskip 0.3 true cm
\noindent
In this appendix we extract the relevant part of the QED cutoff effective 
action using zero-momentum prescriptions (for the case of 
on-shell renormalization prescriptions see for instance \cite{qed}). 
This relevant functional is given by
\formulonaX
{\S_{rel}(\phi;\Lambda,\Lambda_0)&=
\int_k-\frac12A_\mu(-k) (k^2 g^{\mu\nu} Z_A+
g^{\mu\nu}\sigma_2 +k^\mu k^\nu\sigma_\xi)A_\nu(k)\cr
&+\int_p\bar\psi(p)(\slash p Z_\psi-\sigma_m)\psi(p)
+\int_x \sigma_e\bar\psi\slash A\psi+\frac18\sigma_4(A_\mu A^\mu)^2.
}
The relevant couplings are defined by
$$
Z_A=
-\prendi{\partial_{k^2}\frac13 t^{\mu\nu}\S_{\mu\nu}}{k=0},
\quad 
\sigma_2=\prendi{-\frac13 t^{\mu\nu}\S_{\mu\nu}}{k=0},
\quad
\sigma_\xi=
\prendi{\frac1{3}\partial_{k^2}o^{\mu\nu}\S_{\mu\nu}}{k=0},
$$
$$
Z_\psi=\prendi{\partial_{p^\mu}\frac1{16} \gamma^\mu_{\beta\alpha}
\S_{\alpha\beta}}{p=0},
\quad
\sigma_m=
-\prendi{\frac14 \S_{\alpha\alpha}}{p=0},
\quad 
\sigma_e=
\prendi{\frac1{16} \gamma^\mu_{\alpha\beta}\S_{\mu\beta\alpha}}{p_i=0},
$$
$$
\sigma_4=
\prendi{\frac1{72} 
(g^{\mu\nu}g^{\rho\sigma}+g^{\mu\sigma}g^{\nu\rho}+g^{\mu\rho}g^{\sigma\nu})
\S_{\mu\nu\rho\sigma}}{p_i=0}
$$
where $\S_{\mu\nu}$ and $\S_{\alpha\beta}$ are the photon and electron 
two-point functions, $\S_{\mu\alpha\beta}$ and $\S_{\mu\nu\rho\sigma}$ are 
the photon-electron and the four-photon vertices and 
$$
t_{\mu\nu}=g_{\mu\nu}-k_\mu k_\nu/k^2,\quad \quad 
s_{\mu\nu}=g_{\mu\nu}-4k_\mu k_\nu/k^2.  
$$
At the tree level one has
$$
\S^{(0)}=\int_k-\frac12A_\mu (k^2 g_{\mu\nu}+
k^\mu k^\nu(\frac1a-1))A_\nu+\int_p\bar\psi(\slash p-m)\psi
+\int_x e\bar\psi\slash A\psi
$$
\ie
$$
\sigma_A^{(0)}=1,\quad \sigma_\xi^{(0)}=\frac1a-1,\quad\sigma_2^{(0)}=0,
$$
$$
\sigma_\psi^{(0)}=1,\quad\sigma_m^{(0)}=m,\quad\sigma_e^{(0)}=e,\quad
\sigma_4^{(0)}=0.
$$
The irrelevant parts of the photon and electron two-point functions 
are given by
$$
\S_{\mu\nu,irr}(k)=\S_{\mu\nu}(k)
-\bracket{\frac13 t^{\rho\sigma}\S_{\rho\sigma}(\bar k)}{\bar k=0}g_{\mu\nu}
-\bracket{\frac13\partial_{\bar k^2}(t^{\rho\sigma}\S_{\rho\sigma}(\bar k))}
{\bar k=0}k^2g_{\mu\nu}
$$
$$
+\bracket{\partial_{ \bar k^2}s^{\rho\sigma}\S_{\rho\sigma}
(\bar k)}{\bar k=0}k_\mu k_\nu
$$
and 
$$
\S_{\alpha\beta,irr}(p)=\S_{\alpha\beta}(p)-
\prendi{\S_{\alpha\beta}(\bar p)}{\bar p=0}-
p_\mu\prendi{\frac{\partial}{\partial \bar p_\mu}
\S_{\alpha\beta}(\bar p)}{\bar p=0}.
$$
The irrelevant parts of the photon-electron vertex and the four-photon
vertex are given by
$$
\S_{\mu\alpha\beta,irr}(p,p')=\S_{\mu\alpha\beta}(p,p')-
\S_{\mu\alpha\beta}(0,0)
$$
and
$$
\S_{\mu\nu\rho\sigma,irr}(p,q,r)=\S_{\mu\nu\rho\sigma}(p,q,r)-
\S_{\mu\nu\rho\sigma}(0,0,0).
$$ 
The same decomposition into relevant and irrelevant parts holds 
for the vertices of the functional $ I$ and also in the improved case, 
\ie for the functional $\hat\S$ and $\hat I$.

\eject

\end{document}